\documentclass[manuscript]{aastex}

\usepackage{amsmath}
\usepackage{amssymb}
\usepackage{graphicx}
\usepackage{color}
\usepackage{epstopdf}
\usepackage{multirow}

\newcommand{\note}[1]{\emph{\textcolor{red}{}}}

\newcommand{\Sec}{\ensuremath{\mathrm{s}}}
\newcommand{\yr}{\ensuremath{\mathrm{yr}}}
\newcommand{\kpc}{\ensuremath{\mathrm{kpc}}}
\newcommand{\kms}{\ensuremath{\mathrm{km}\,\mathrm{s}^{-1}}}
\newcommand{\Myr}{\ensuremath{\mathrm{Myr}}}
\newcommand{\K}{\ensuremath{\mathrm{K}}}
\newcommand{\B}{\ensuremath{\mathrm{B}}}
\newcommand{\eV}{\ensuremath{\mathrm{e\!V}}}
\newcommand{\gcc}{\ensuremath{\mathrm{g}\,\mathrm{cm}^{-3}}}

\newcommand{\HI}{H~I}
\newcommand{\HeI}{He~I}
\newcommand{\HII}{H~II}
\newcommand{\HeII}{He~II}
\newcommand{\HeIII}{He~III}

\newcommand{\cc}{\ensuremath{\mathrm{cm}^{-3}}}

\newcommand{\foe}{\ensuremath{\mathrm{B}}}

\newcommand{\cm}{{\ensuremath{\mathrm{cm}}}}

\newcommand{\usec}{{\ensuremath{\mathrm{sec}}}}
\newcommand{\pc}{{\ensuremath{\mathrm{pc}}}}
\newcommand{\erg}{{\ensuremath{\mathrm{erg}}}}

\newcommand{\Mpc}{{\ensuremath{\mathrm{Mpc}}}}

\newcommand{\Hm}{{\ensuremath{^{} \mathrm{H}_2}}}

\newcommand\Msun{{\rm M_\odot}}

\newcommand\Zsun{{\rm \,Z_\odot}}
\newcommand{\unit}[1]{\ensuremath{\, \mathrm{#1}}}

\newcommand{\GADGET}{\texttt{GADGET}}

\renewcommand{\thefootnote}{\ifcase\value{footnote}\or*\or
(**)\or(***)\or(****)\or(\#)\or(\#\#)\or(\#\#\#)\or(\#\#\#\#)\or($\infty$)\fi}

\makeatletter

\newcommand{\Rmnum}[1]{\expandafter\@slowromancap\romannumeral #1@}
\makeatother

\begin{document}
\title{Cosmological Impact of Population~III Binaries} 

\author{ Ke-Jung Chen\altaffilmark{1,2,*}, Volker Bromm\altaffilmark{3}, 
Alexander Heger\altaffilmark{2,4,5},  Myoungwon Jeon\altaffilmark{3}, and Stan  Woosley\altaffilmark{1} } 

\altaffiltext{1}{Department of Astronomy \& Astrophysics, University of California, Santa 
Cruz, CA 95064, USA} 

\altaffiltext{2}{School of Physics and Astronomy, University of Minnesota, Minneapolis, MN 
55455, USA}

\altaffiltext{3}{Department of Astronomy, University of Texas, Austin, TX 78712, USA } 

\altaffiltext{4}{Monash Centre for Astrophysics, School of Mathematical Sciences, Monash University, Victoria 3800, Australia} 

\altaffiltext{5}{Joint Institute for Nuclear Astrophysics, University of Notre Dame Notre Dame, IN 46556 USA} 

\altaffiltext{*}{IAU Gruber Fellow; kchen@ucolick.org}

\begin{abstract}
We present the results of the stellar feedback from Pop~III binaries by employing 
improved, more realistic Pop~III evolutionary stellar models. To facilitate a meaningful 
comparison, we consider a fixed mass of $60\,\Msun$ incorporated in Pop~III stars, 
either contained in a single star, or split up in binary stars of $30\,\Msun$ each or
an asymmetric case of one $45\,\Msun$ and one $15\,\Msun$ star. Whereas the sizes of the resulting 
\HII{} regions are comparable across all cases, the \HeIII{} regions around binary stars are 
significantly smaller than that of the single star. Consequently, the He$^{+}$ 1640 $\textup{\AA}$ 
recombination line is expected to become much weaker. Supernova feedback exhibits great 
variety due to the uncertainty in possible explosion pathways. If at least one of the component 
stars dies as a hypernova about ten times more energetic than conventional core-collapse 
supernovae, the gas inside the host minihalo is effectively blown out, chemically enriching the 
intergalactic medium (IGM) to an average metallicity of $10^{-4}-10^{-3}\,\Zsun$, out to $\sim 2\,\kpc$. 
The single star, however, is more likely to collapse into a black hole, accompanied by at most very weak explosions.
The effectiveness of early chemical enrichment would thus be significantly reduced, in
difference from the lower mass binary stars, where at least one component is likely
to contribute to heavy element production and dispersal. Important new feedback physics is also
introduced if close binaries can form high-mass x-ray binaries, leading to the pre-heating and -ionization
of the IGM beyond the extent of the stellar \HII{} regions.
\end{abstract}


\section{Introduction}
One of the paramount problems in modern cosmology is to elucidate how the first 
generation of luminous objects, stars, accreting black holes (BHs) and galaxies, shaped the early universe 
at the end of the cosmic dark ages \citep{barkana2001, loeb2013, wiklind2013}.
A key driver of this grand cosmic transformation was the gradual enrichment of the pristine
universe with heavy chemical elements in the wake of the first supernova (SN) explosions
 \citep[reviewed in][]{karlsson2013}. According to the modern theory of cosmological structure
formation \cite[e.g.][]{mo2010}, the hierarchical assembly of dark matter (DM) halos provided 
the gravitational potential wells that allowed gas to form stars and galaxies inside them. 
Extending this model to the highest redshifts, one can determine the sites where the first
stars, the so-called Population~III (Pop~III), formed out of the pure H/He gas created in
the Big Bang. Within this framework,
Pop~III stars are predicted to form inside
of DM minihalos with total masses (DM and gas) of about $10^{6}\,\Msun$ at redshifts
of $z\sim 20 - 30$ \citep{couchman1986, haiman1996, tegmark1997, bromm2004, oshea2008, bromm2009}.

The first stars affected the early universe in several different ways.
Massive Pop~III stars were strong emitters of hydrogen and helium ionizing photons that built 
up extensive {\HII}, {\HeII}, and {\HeIII} regions \citep{yoshida2007a}. The metals forged in Pop~III stars 
later were dispersed into the intergalactic medium (IGM) when they died as SNe, thus quickly polluting the primordial gas such that the second generation of (Population~II) stars could emerge.
It is convenient to classify Pop~III feedback mechanisms into different classes (Ciardi \& Ferrara
2005), specifically radiative feedback
\citep{oet05, abel2007, su07, wet08b, hus09, wet10, yoshida2007a, greif2009a}, mechanical and chemical feedback due to SNe \citep{wise2008, greif2010, ritter2012}, 
and X-ray feedback from accreting BH remnants \citep{kuhlen2005,alvarez2009, jeon2012, jeon2014, xu2014}.

In the Pop~III star-forming regions, metal
cooling was absent because the primordial gas consisted almost exclusively of
hydrogen ($\sim76\,\%$ by mass) and helium ($\sim24\,\%$). Molecular hydrogen was
thus the dominant coolant, but owing to its quantum-mechanical structure,
it was unable to cool the gas to the low temperatures typically encountered in star forming
clouds today. The primordial gas, therefore, remained relatively warm, with
typical temperatures of several hundred Kelvin. Hence the Jeans mass was correspondingly larger, as
well, leading to the expectation that Pop~III stars 
might have been more massive than stars formed today, with a predicted mass scale of
$50\,\Msun - 100\,\Msun$ \citep{bromm1999, bromm2002, naka2001, abel2002, omukai2003}.
Because of their high surface temperatures
\citep{bond1984, bkl2001, schaerer2002}, Pop~III stars could effectively
produce copious amounts of ionizing UV photons.
Extended \HII{} regions with size of several
kpc were created before the stars died. Given
the shallowness of the gravitational potential well of the host DM halos, the
surrounding gas was subject to strong photo-heating, thus being able to easily escape the
host minihalos \citep{whalen2004, alvarez2006}. This photo-evaporation suppressed
further star formation inside the minihalos, thus delaying further star formation until
more massive host halos emerged \citep{bromm2011}.

The character of Pop~III feedback sensitively depends on the fate encountered
by massive Pop~III stars when they die after their
short lifetime of a few million years to trigger a SN explosion,
or directly collapse into
black holes. Those Pop~III stars with masses of $140\,\Msun$ -
$260\,\Msun$ are thought to die as pair-instability supernovae (PSNe)
\citep{barkat1967, heger2002, chen2014}, although this mass range may have to be revised in the
case of rapidly rotating progenitors \citep{chatz2012}.
Unlike gravitationally-powered core-collapse supernovae
(CCSNe), PSNe are hyper-energetic thermonuclear explosions, not leaving any
compact remnant behind. Because vast amounts of metals
are ejected during a PSN explosion, even a single event could
enrich about $10^7\,\Msun$ of primordial gas up to
$10^{-4}\Zsun-10^{-3}\Zsun$ \citep{karlsson2008,wise2008,ss09,greif2010}.
Even such a trace amount of metals could change the subsequent star
formation process and might cause a transition of the stellar initial mass
function (IMF) from the top-heavy mode predicted for
Pop~III stars to the standard IMF for later (Pop~I and Pop~II)
generations with typical masses comparable to that of our Sun
\citep{bfcl2001, omukai2005, maio2010, wise2012}.

Results from stellar archaeology \citep{beers2005, frebel2005}, which
studies the most metal-poor stars in the Local Group that retained the imprints from
nucleosynthesis in the early universe, possibly including those from Pop~III stars, in general
do not support the chemical abundance pattern predicted for PSN enrichment \citep{tumlinson2006}. Theoretical PSN
yields exhibit a pronounced odd-even effect resulting from a low
neutron excess \citep{heger2002}. In addition,
the lack of any neutron capture process results in the absence of all elements heavier
than the Fe peak (no r- or s-process).
The Pop~III CCSN
models \citep{umeda2003, heger2010}, however, can produce abundance
patterns in good accord with the observation of metal poor stars.
Recent simulations of Pop~III star formation that take into account the radiation-hydrodynamical
feedback from the growing central protostar have shown that
accretion can be halted, thus preventing
the formation of stars more massive
than 50 $\Msun$ \citep{hoso2011, stacy2012, hirano2014}. That implies that Pop~III stars typically might
die in a CCSN, instead of a PSN, in agreement with the observations.
Furthermore,
recent cosmological simulations with extremely high resolution have shown
that the primordial gas cloud is able to fragment and produce stars
of relatively lower mass of tens of solar masses \citep{turk2009, stacy2010, greif2011}, 
organized in binaries or multiple stellar
systems. These simulations suggest that binary systems may be the typical
channel for primordial star formation in minihalos \citep{stacyB2013}.
Since these simulations only follow the protostellar assembly process for at most
$\sim 10^3$\,yr, it is not yet clear how the final mass spectrum will look like \citep{bromm2013}.
One key uncertainty is the degree of merging of neighboring protostars \citep{greif2012}.
However, it appears likely that the first stars in minihalos typically formed in binary
or small multiple systems.
Since the evolution of binary systems
and their final fate are very different from those of single stars, it is
worthwhile to investigate whether binary Pop~III stars lead to significantly altered feedback effects.

Since their evolution is quite
different from a single
star, it is worth investigating how the Pop~III binary
systems affected the IGM and their host halos 
that later merged into the first galaxies \citep{rgs02,wise2008,wise2009,wise2014,johnson2009,johnson2013,xu2013}. 
The evolving binaries might exert different feedback
mechanisms, through the emission of UV and x-ray ionizing photons, and SN
explosions, all of which may be quite different from the feedback 
of the single stars which has been well documented in the literature. Therefore,
we first study the impact of the first massive stars of  masses $60\,\Msun$, 
$45\,\Msun$, $30\,\Msun$, and $15\,\Msun$ on their parent halos. 
 
 We study the possible impacts of the first binary systems on the IGM and 
present the results of cosmological simulations by considering possible outcomes
of the Pop~III binary models with stars of $45\,\Msun$+$15\,\Msun$ (S45+S15)
and $30\,\Msun$+$30\,\Msun$ (S30+S30). Our binary models consider the
non-interacting binaries during their stellar evolution. However, more
realistic binary models might have a much wider range of outcomes
\citep{langer2012}.

The structure of this paper is as follows: In Section~2, we
describe our initial setup, as well as our numerical methods. A discussion of our 
protostellar evolution models, both for single and binary stars, follows in Section~3.
The simulation results are presented in Section~\ref{results}, and we conclude
by discussing the broader implications
in Section~\ref{flms_discussion}.  All of the results presented in this paper use physical 
coordinates instead of comoving coordinates.

\section{Numerical Methodology}
\subsection{Problem Setup}
\label{flms_nm} 
The primary code used for our simulations is the well-tested,
massively-parallel cosmological code \GADGET{} \citep{springel2005},
which computes gravitational forces with a hierarchical tree algorithm
and represents fluids by means of smoothed particle hydrodynamics
(SPH). In order to simulate the feedback exerted by the first stars,
additional physical processes, such as cooling and chemistry of the primordial gas, radiative
transfer of ionizing photons, and SN explosions, are
required and have been implemented in \GADGET{}.   

Our simulations employ the same initial conditions as in \cite{greif2010},
starting at $z = 100$ in a periodic box of linear size of $1\,\Mpc$ (co-moving). We choose
 $\Lambda$CDM cosmological parameters with matter density $\Omega_m=0.3$, baryon density $\Omega_b=0.04$, 
 Hubble constant $H_0 = 70\,\unit{km}$\,$\unit{s^{-1} Mpc^{-1}}$, spectral index $n_{\rm s}=1.0$, and normalization 
 $\sigma_8=0.9$, based on the WMAP cosmic microwave background (CMB) measurement \citep{komatsu2009}. 
\cite{greif2010} used a standard hierarchical zoom-in technique, generating the highest mass
resolution covering the Lagrangian region where the first galaxy is destined to form.
That way, all the relevant fine-structure, specifically the minihalo progenitors
of the first galaxy, can be resolved. The resulting mass of the DM and gas particles
in the highest resolution region is $m_{\rm DM}\sim33\,\Msun$ and $m_{\rm sph}\sim5\,\Msun$, 
 respectively. Because the molecular hydrogen cooling in primordial gas imprints a characteristic 
 density of $n_{\rm H}=10^4\,\unit{cm^{-3}}$ and temperature of $200\,\K$ \citep{bromm2002}, our simulations marginally resolve the corresponding Jeans mass of $M_{\rm J}\simeq 500 \Msun$. 
Our cooling module and chemistry network are based on \cite{greif2010},  
and include all relevant cooling mechanisms of primordial gas, such as H and He collisional 
ionization, excitation and recombination cooling, bremsstrahlung, 
and inverse Compton cooling; in addition, the collisional excitation cooling via $\Hm$ and HD is also taken into account. For $\Hm$ cooling, 
collisions with protons and electrons are explicitly included. The chemical network includes $\rm H, H^+, H^-, H_2, H_2^+, He, He^+, He^{++}$, and $\unit{e^-}$, D, $\rm D^+$, and HD.

State-of-the-art cosmological simulations can potentially use billions of particles
to model the formation of the universe. However, it is still extremely challenging 
to resolve mass scales from galaxies (10$^{10}\,\Msun$) to individual
stars ($1\,\Msun$). For example, the resolution length in our 
simulation is about $1\,\pc$, many orders of magnitude removed from truly stellar scales.
Hence, modeling the process of star formation on cosmological scales from first 
principles is currently still out of reach. A viable strategy to treat star 
formation and its feedback is to employ sub-grid models, where sink particles approximately
represent unresolved single stars, or clusters thereof. The sinks can then act as
sources of radiation, with luminosities and spectra chosen in accordance with
stellar structure and evolution models, and eventually as sites for SN explosions.
Another reason to use sink particles is to overcome the so-called `Courant myopia'.
When the gas density somewhere inside the computational box becomes increasingly high, 
the SPH smoothing length decreases, in turn enforcing the adoption of smaller and smaller timesteps, according to the Courant condition. When encountering a runaway collapse, the simulation
would effectively grind to a halt or fail.
Creating sink particles allows us to bypass this numerical bottleneck, such that the 
simulations can be followed beyond the initial collapse, where much of the interesting
physics, related to the stellar feedback, occurs.
We here apply the sink particle algorithm of \cite{JB2007}.
The key criterion for sink creation,
and subsequent accretion of further SPH particles, is that the gas density exceeds a
pre-specified density threshold,
$n_c\,\sim\,10^4\,\cm^{-3}$, but we also test for gravitational boundedness, and
whether the gas is part of a converging flow, where
$\nabla\cdot\vec{v}<0$.  
The sink particles 
provide markers for the position of a Pop~III star and its remnants, such as a black hole or SN, 
to which detailed sub-grid physics can be supplied. 
 
 \subsection{Radiative Transfer}
 \label{ray_sec}
 When a Pop~III star has formed inside the minihalo, the sink particle representing it
 immediately turns into a point source of ionizing photons to mimic the
 birth of a star. The rate of ionizing photons emitted depends on the
 physical size of the star and its surface temperature based on our stellar subgrid 
 models. Instead of simply assuming constant rates of emission, 
 we use the results of one-dimensional stellar evolution calculations from \cite{heger2010} 
 to construct the luminosity history of Pop~III stars. 
 Indeed, luminosities exhibit considerable time variability when taking the
 evolution off the main sequence into account.
 The photons streaming from the star then establish an ionization front and build 
 up \HII{} regions. To trace the propagation of photons and the ionization front, 
 we use the ray-tracing algorithm from \citep{greif2009a},  which solves the
 ionization front equation in a spherical grid by tracking $10^5$ rays
 with 500 logarithmically spaced radial bins around the photon source.
 The radiation transport is coupled to the hydrodynamics of the gas through 
 its chemical and thermal evolution. The transfer of the $\rm H_2$-dissociating photons
 in the Lyman--Werner (LW) band ($11.2-13.6\,\eV$) is also 
 included.
 
 For completeness, we here briefly present the key features of the ray-tracing algorithm,
 and refer the reader to \cite{greif2009a} for details and tests.
 To begin with, particle positions are transformed from Cartesian to 
 spherical coordinates, i.e. radius ($r$), zenith angle ($\theta$), and azimuth angle ($\phi$). 
 The effective volume of each particle is  $\sim h^3$, where $h$ is the SPH
 smoothing length.  The corresponding sizes in spherical coordinates are $\Delta r\,=\,h$,
  $\Delta \theta\,=\,h/r$, and  $\Delta \phi\,=\,h/r\sin\theta$. Using spherical coordinates  
 facilitates the convenient calculation of the
 ionization front around the star, 
 \begin{equation}
 n_n r^2_{\rm I}\frac{d r_{\rm I}}{dt}\,=\, \frac{\dot{N}_{\rm ion}}{4\pi}\,-\,\alpha_{\rm B}\int_0^{r_{\rm I}}n_en_+r^2dr, 
 \label{rad1}
 \end{equation}
 where $r_{\rm I}$ is the position of the ionization front, $\dot{N}_{\rm ion}$ represents 
 the number of ionizing photons emitted from the star per second, $\alpha_{\rm B}$ is
 the case B recombination coefficient, and $n_n$, $n_e$, and $n_+$ are the number densities of neutral
 particles, electrons, and positively charged ions, respectively. The recombination coefficient 
 is assumed to be constant at temperatures around $2\times10^4\,\K$. The ionizing photon rates
 are
 \begin{equation}
 \dot{N}_{\rm ion}\,=\,\frac{\pi L_*}{\sigma_{\rm SB} T_{\rm eff}^4}\,\int_{\nu_{\rm min}}^{\infty}\frac{B_{\nu}}{h_{\rm P}\nu} {\rm d}\nu, 
 \label{rad2}
 \end{equation}
 where $h_{\rm P}$ is Planck's constant, $\sigma_{\rm SB}$ the Stefan-Boltzmann constant, $\sigma_{\nu}$ the relevant photo-ionization 
 cross section, and $\nu_{\rm min}$ the ionization threshold for
 \HI, \HeI, and  \HeII. 
 By assuming a blackbody spectrum with effective temperature, $T_{\rm eff}$,
 the flux of a Pop~III star can be written 
 \begin{equation}
 F_{\nu}\,=\,\frac{L_*}{4 \sigma T_{\rm eff}^4r^2}B_{\nu}. 
 \end{equation}
 The size of the \HII{} region is determined by solving Equation~\ref{rad1}. 
 The particles within the \HII{} regions store information
 about their distance from the star, which is used to calculate the 
 ionization and heating rates,
 
 \begin{displaymath}
 k_{\rm ion}\,=\,\int_{\nu_{\rm min}}^{\infty}\frac{F_{\nu}\sigma_{\nu}}{h_{\rm P}\nu} {\rm d}\nu,
 \end{displaymath}

 \begin{equation}
 \Gamma\,=\,n_n\int_{\nu_{\rm min}}^{\infty}F_{\nu}\sigma_{\nu}\Big(1\,-\,\frac{\nu_{\rm min}}{\nu}\Big){\rm d}\nu. \
 \end{equation}
 
 { \Hm{} is the most important coolant for cooling the 
 primordial gas, which leads to formation of the first stars. However, its hydrogen bond is weak 
 and can be easily broken by photons in the LW 
 bands between 11.2 and 13.6 eV. The small $\Hm$ fraction in the IGM 
 creates only a little optical depth for LW photons, allowing them to propagate a much
 larger distance than ionizing photons. In our algorithm, self-shielding of H$_2$ is not 
 included because it is only important when H$_2$ column densities are high. 
 Here we treat the photodissociation of $\Hm$ in the optically thin limit and the dissociation rate in a 
 volume constrained by causality  within a radius, $r =ct $. The dissociation rate 
 is given by $k_{\rm H_2}=1.1\times10^8 F_{\rm LW}\,\usec^{-1}$, where $F_{\rm LW}$
 is the flux within LW bands \citep{greif2009a}.}

 \subsection{X-Ray Emission}
 \label{xray_sec}
 A compact binary may be able to produce
 radiative feedback in the form of x-rays. In this section, we 
 describe the treatment of the radiation from such an x-ray binary source, if present. 
 Our methodology is based on \cite{jeon2012}. In the 
 local universe, the non-thermal emission spectrum from accreting back holes or neutron
 stars can be expressed as $F_{\nu}\propto \nu^{\beta}$, where
 $\beta = -1$ is the spectral index. {  More precisely, the full spectrum is a combination 
 of a power-law component, representing non-thermal synchrotron radiation with the luminosity 
 formula $L \sim \dot{M}c^2$, where $\dot{M}$ is the Bondi-Holye accretion model \citep{bondi1944}, 
 and a thermal multi-color disk component}. We here ignore the latter, as we are only interested 
 in the x-ray feedback on the general IGM, where only the optically-thin, non-thermal photons
 contribute.
 We conservatively assume that the BH emission physics in the early universe is identical to the local case, 
 and we therefore apply the same spectra for the black holes originating from the first stars.  
 
 The propagation of high-energy photons is assumed to result in an isotropic radiation field, 
 $\propto 1/r^2$, which only depends on the distance from the BH.  
 The corresponding photo-ionization and photo-heating rates can be 
 written as \citep{jeon2012}: 
 \begin{equation}
 \label{ion1}
         k_{\rm ion} = \frac{\dot{K}}{r^2_{\unit{pc}}} \left(\frac{\dot{M}_{\rm BH}}{10^{-6}
      \Msun\,\yr^{-1}}\right),
 \end{equation}
 where
 \begin{equation}
 	\dot{K} = [  1.96, 2.48, 0.49  ] \times 10^{-11} \hspace{0.5cm} \quad\Sec^{-1} .
 \end{equation}
 and
 \begin{equation}
 \label{heat1}
         \Gamma= \frac{n_{j}\dot{H}}{r^2_{\unit{pc}}}
  \left(\frac{\dot{M}_{\rm BH}}{10^{-6}\Msun\,\rm yr^{-1}}\right) \left(1-f_{\rm H/He}\right) \hspace{0.1cm},
 \end{equation}
 Here
 \begin{equation}
         \dot{H} = [  7.81, 9.43, 1.63  ] \times 10^{-21} \hspace{0.3cm} \quad\erg\,\cc\,\Sec^{-1}
 \end{equation}
 for \HI, \HeI, \HeII, respectively. $n_{j}$ is the corresponding number density,
 $r_\pc$ the distance from the star in units of $\pc$, and $\dot{M}_{\rm BH}$ the mass accretion rate. Finally, $f_{\rm H/He}$ are the fractions of the total photon energy expended in 
 secondary ionizations \citep{shull1985}.
 
 \subsection{Supernova Explosion and Metal Diffusion}
 \label{sn_sec}
 After several million years, the massive Pop~III
 stars eventually exhaust their fuel, and many of
 them might have died as supernovae or black holes. 
 As we discussed above, the first SN explosions may be extremely powerful,
 accompanied by huge outputs of energy and metals. 
 Here, we briefly discuss how we model SN explosions in 
 our cosmological simulations. 
 
 When the star reaches the end of its lifetime, we initialize
 a SN blast wave by
 distributing the explosion energy among the SPH particles surrounding
 the sink that had marked the location of the Pop~III star. Because the resolution of the simulation 
 is about 1~pc, we cannot resolve individual 
 SNe in both mass and space. Instead, we here select the particles within a region 
 of 10~pc to share the supernova's thermal energy and metal yield. The gas within 
 this region has mean temperatures about several million Kelvin. On this scale, the blast 
 wave is still close to its energy-conserving phase. The explosion energy of hypernovae 
 and pair-instability SNe can be up to $10^{52}-10^{53}\,\erg$, whereas a conventional
 core-collapse SN has about $10^{51}\,\erg$. 
 
 Mixing plays a crucial role in the transport of metals, which could be the most 
 important coolant for subsequent star formation. We here cannot resolve the fine-grained
 mixing due to
 fluid instabilities in the early SN ejecta, developing on a scale far below 1~pc. 
 However, we approximately model the coarse-grain mixing on cosmologically relevant
 timescales by applying the
 SPH diffusion scheme from \cite{greif2009b}.
 A precise treatment of the mixing of metals in cosmological simulations 
 is not available so far because the turbulent motions responsible for 
 mixing can cascade down to very small scales, far beyond the resolutions 
 we can achieve now. We here, therefore, approximately model the effect of
 unresolved, sub-grid turbulence as a diffusion process, linking the corresponding
 transport coefficients to the local physical conditions at the grid scale.
 For further algorithmic details, we refer the reader to \cite{greif2009b}.
 
 After the SN explosion, metal cooling must be considered in the cooling network. We assume that C, O, 
 and Si are produced with solar relative abundances, which are the dominant coolants for the gas contaminated 
 by the first SNe. 
 There are two distinct temperature regimes for these species. In low temperature gas, 
  $T\,<\,2\,\times\,10^4\,\K$, we use a chemical network presented in \cite{glover2007}, which follows 
  the chemistry of C, C$^+$, O, O$^+$, Si, Si$^+$, and Si$^{++}$, supplemental to the primordial species discussed above. 
  This module
  considers the fine structure cooling of C, C$^+$, O, Si, and Si$^+$, whereas molecular 
  cooling is not taken into account. At high temperatures, 
 $T\,\geq\,2\,\times\,10^4\,\K$, due to the increasing number of ionization states, a full non-equilibrium 
 treatment of metal chemistry becomes computationally prohibitive. Instead of directly 
 solving the cooling network, we use the cooling rate table based on \cite{suther1993}, which gives effective 
 cooling rates for hydrogen and helium line cooling, as well as bremsstrahlung, at different metallicities.  
 Dust cooling is not included because it would only become important at densities much higher than what is
 reached in our simulations.

\section{Stellar Models}

\subsection{Single Star Models}
We use the Pop~III stellar models of $15\,\Msun$ (S15), $30\,\Msun$ (S30), 
$45\,\Msun$ (S45), and $60\,\Msun$ (S60) stars from the library of 
\citet{heger2010}. These models are non-rotating stars and we assume 
no mass loss during the stellar evolution. We summarize key model characteristics 
in Table~\ref{flms_1ife_models}, where we distinguish the life 
span of main-sequence (MS; central hydrogen burning) and post-MS 
(until supernova) evolution. The S15 model evolves in total for about $10.5\,\Myr$ 
before encountering an iron core-collapse supernova with an explosion 
energy of $1.2\times10^{51}\,\erg =1.2\,\B$ and a metal yield of $1.4\,\Msun$. 
Similarly, the S30, S45, and S60 models evolve for $5.7$,  $4.4$, and $3.7\,\Myr$, respectively. 
In assigning the final fate of our three most massive models, current understanding is
still quite uncertain, and we here focus on a few illustrative possibilities.
Specifically, in the cases of 
stars with masses of $30\,\Msun$ and above, we assume that they do not die as 
conventional iron core-collapse SNe. Instead, we assume that each of them either collapses 
into a black hole (BH), triggering no explosion, or explodes as a hypernova (HN) with 
$10\,\B$ explosion energy, based on the collapsar model \citep{woosley1993}. 
When the star dies as a BH, all metals within the star fall back into the BH, such
that no enriched material is ejected.
In the HN case, the S30, S45, and S60 models synthesized about $6.8$,  $13.2$, 
and $20.6\,\Msun$ of heavy elements, and disperse them into the primordial IGM. We summarize
the possible stellar fates in Table~\ref{flms_fate}. For simplicity and to limit the 
number of cases in this study, we focus only on these simplified, limiting cases, and 
refer the reader to \citet{heger2010} for an extended discussion
of Pop~III SN models. 

\begin{figure}[h]
\begin{center}
\includegraphics[width=.6\columnwidth]{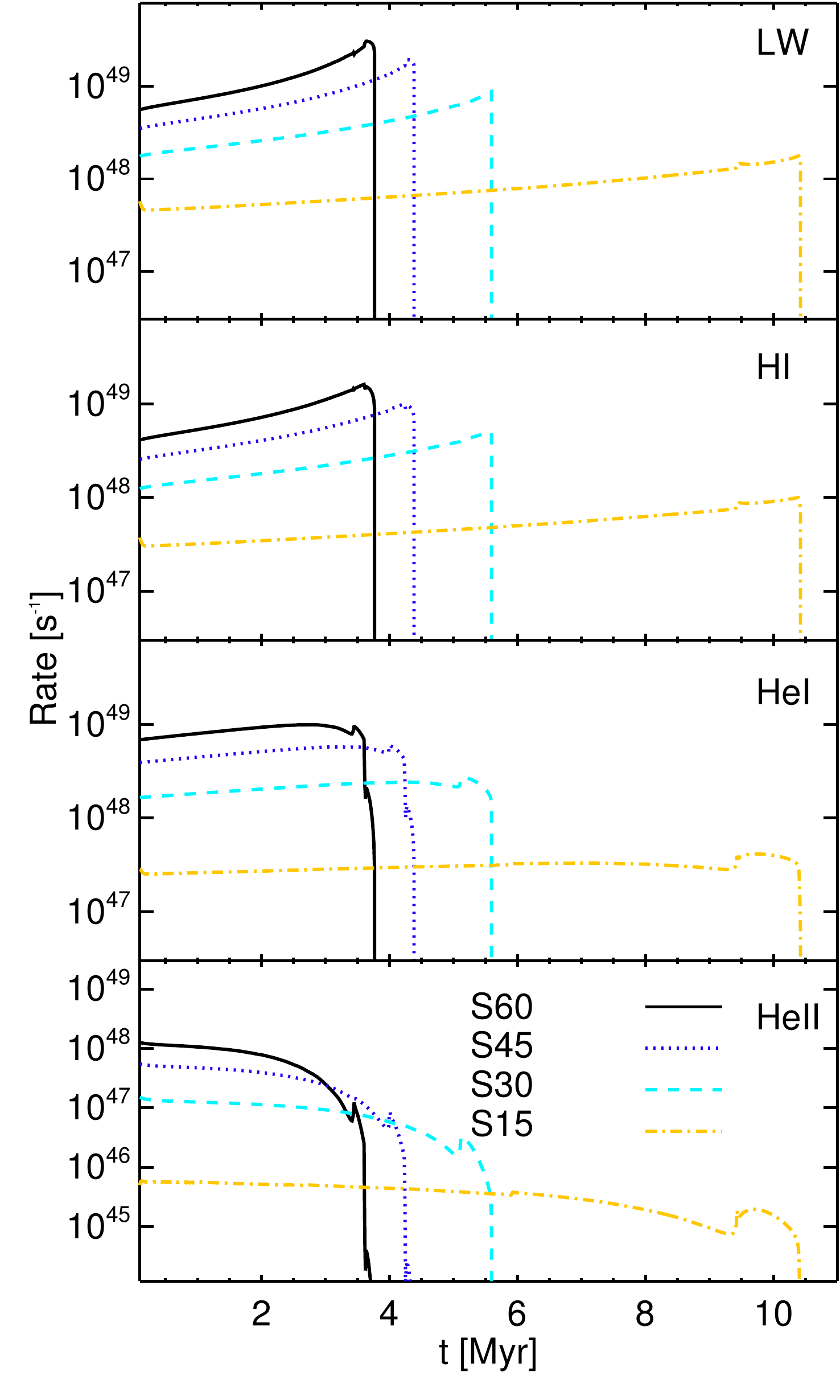} 
\end{center}
\caption[Evolution of the ionizing photon rate]{Evolution of UV flux of 
$60\,\Msun$, $45\,\Msun$, $30\,\Msun$, and $15\,\Msun$ Pop~III stars. During
the main sequence stage, rates are close to constant. After leaving the 
main sequence, the luminosity of the stars increases due to the expansion of the envelope, 
leading a drop in temperature. This leads to an increase in the \HI{} ionizing photon rate
 (lower energy band), but a decrease in the rate of {\HeII} ionizing photons 
 (higher energy band).\label{flms_photons}}.
 \end{figure}

Massive Pop~III stars are strong sources of UV radiation. Because of the 
predicted weak stellar 
winds from Pop~III stars \citep{kudri2002}, there is no notable x-ray source contributing 
to the ionizing photon budget resulting from such winds. The UV radiation thus exclusively
emerges from the hot stellar surface. Ionizing photon fluxes for all stellar models are
given in Figure~\ref{flms_photons}, where evolutionary effects are evident. Specifically,
fluxes exhibit a gradual increase in the lower energy bands  
(LW and \HI), and a decrease in the higher energy band (\HeII). The hydrogen ionizing flux 
in the S60 model is about 10 times larger than for S15. The flux of more energetic 
photons is highly sensitive to stellar mass, such that for photons capable of ionizing
\HeII{}, their ratio is 100:10:1 for S60:S30:S15. Because the life span 
of each star is different, we evaluate the overall ionizing power of an individual star 
by calculating the total amount of ionizing photons emitted before the star dies. 
As shown in Table~\ref{flms_flux}, the S60 model produces about 2, 3, and 6 times more
\HI{}, \HeI{}, \HeII{} ionizing photons than S30. The cumulative ionizing power of S60 is also much 
stronger than the production from four S15 models combined. This implies that the overall radiative feedback 
of Pop~III stars becomes weaker if their mass scale shrinks due to fragmentation. 

Since our ray-tracing scheme cannot resolve a time scale less than a year, the 
radiation flash from the SN itself is not included here because the SN transit only lasts for 
about a few weeks to months. In principle, there could be a flash 
of hard radiation from the shock breakout that may eventually be observable 
\citep{scannapieco2005}, but the total energy in this flash is small, due largely to 
the typically small radii of the Pop~III stars at the time of death.  
Besides, the radiation from the subsequent main part of the supernova light 
curve is largely at longer wavelengths and does not contribute much to the ionization.
Our calculation shows that the total ionizing photon production during the SN is about $10^{-5}$ 
that of the MS phase. Furthermore, most of the SN explosion energy goes 
into the thermal and kinetic energy of the ejecta \citep{souza2013, whalen2014a}.  

\subsection{Binary Star Models}
The ubiquitous fragmentation of primordial star forming clouds allows the widespread
formation of binary stars. We consider binary stars with a total mass 
of $60\,\Msun$, specifically a system with two stars of equal mass (mass ratio 1:1)
and another one with a mass ratio of 3:1, in accordance with current theoretical understanding.
\citet{shu1987} and \citet{larson2003} suggest that binary stars with (close to) 
equal mass are common in the local universe. Our binary models thus contain both asymmetric
cases of $45\,\Msun + 15\,\Msun$ Pop~III stars and symmetric ones of 
$30\,\Msun + 30\,\Msun$ Pop~III stars.
To keep the binary models simple, we do not consider binary star mergers, and neglect
any mass ejection as an idealized approximation. We are aware of and advise the 
reader of the shortcomings of these simplifications, compared to more realistic binary 
models \citep{langer2012}. 
Below, we discuss the select binary scenarios considered here (see Table~\ref{binaryfate}).

\begin{enumerate}
\item{\bf S30+S30 (HN)}: 
This model contains two $30\,\Msun$ Pop~III stars, each represented by our S30 model.
We assume both stars form at the same time 
and evolve together for about $5.7\,\Myr$, after which they both die as a HN.

\item {\bf S30+S30 (BH)}: 
This scenario is very similar to S30+S30 (HN), but now both stars
directly collapse into BHs, without triggering a SN.

\item {\bf S45+S15 (BH)}:  
The binary contains two stars, represented by S45 and S15 models. Both of them form at the 
same time and evolve together for $4.4\,\Myr$. Subsequently, the S45 component dies as a BH,
and the system becomes an x-ray binary source due to the transfer of mass
from the S15 companion onto the BH. We approximately assume that the entire mass of the primary 
collapses into the BH, and that the system remains bound.  
We employ a mass transfer rate of $10^{-6}\,\Msun$/yr, active for about
$10\,\Myr$. Because the S15 secondary would lose much of its mass during this x-ray binary phase, 
we assume a white dwarf (WD) death, without a SN explosion.

\item {\bf S45+S15 (HN)}: Both stars again form at the same time 
and evolve together for $4.4\,\Myr$, then the S45 primary dies as a 
HN. The S15 component evolves for another $6\,\Myr$, then 
dies as a CCSN. For a wide binary, the kick 
velocity is small, and the star will die essentially at the 
location of the original binary. However, in a close binary situation, 
the orbital velocity is high, and the S15 secondary would acquire a kick of about
$100\,\kms$, assuming a binary separation of about 1~AU.
 Such a velocity might allow the star to travel close to 
1~kpc before dying as a SN. In this case, the ejected S15 model could
become a moving radiation source and disperse its metal production from a site
quite remote from where it formed. \citet{conroy2012} suggested that the 
runaway massive stars could also contribute to the reionization of the Universe. 
For simplicity, we do not explore this intriguing scenario in this paper.
\end{enumerate}

\begin{figure}[h]
\begin{center}
\includegraphics[width=.6\columnwidth]{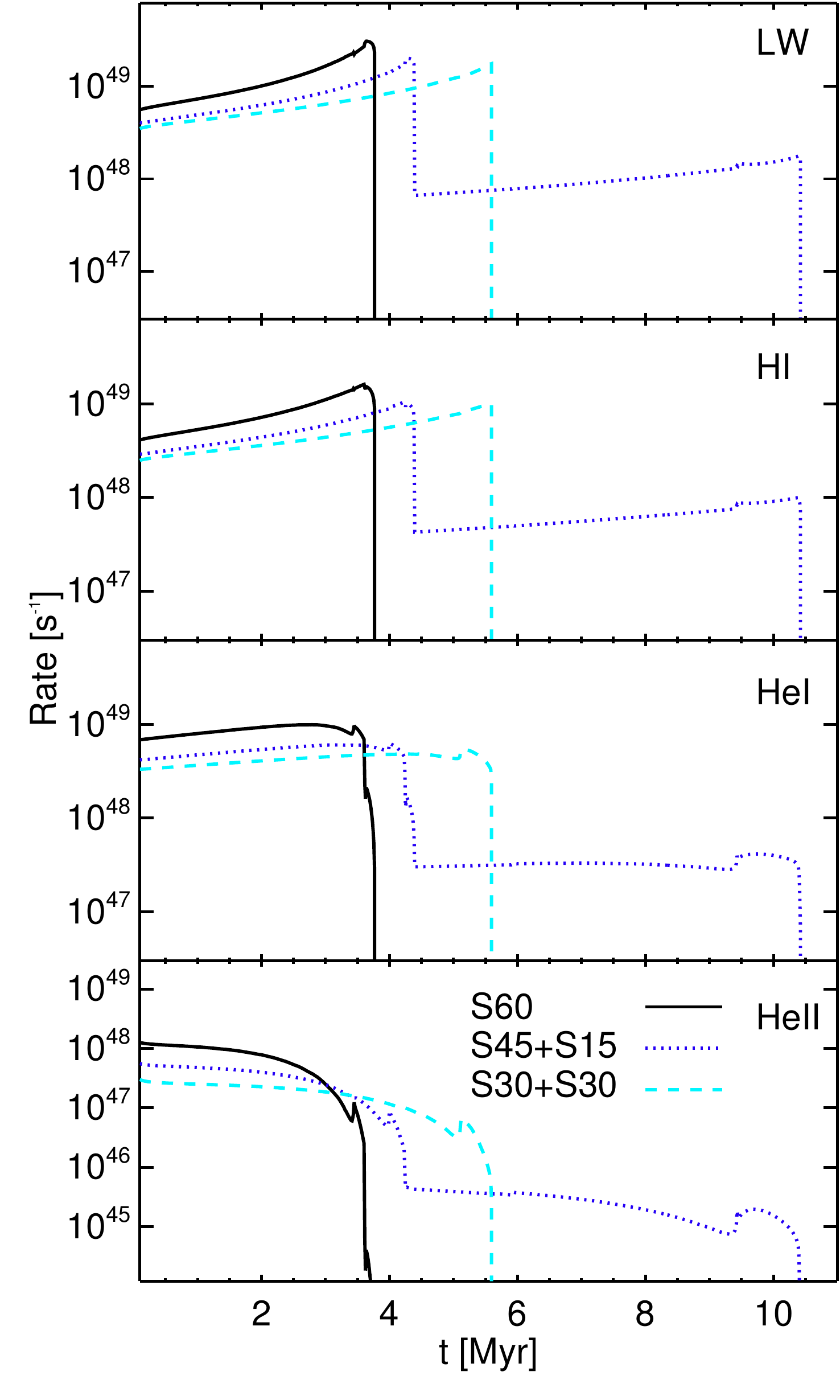} 
\end{center}
\caption[Evolution of the ionizing photon rate]{UV fluxes of the
first binary stars. Note that fluxes are calculated by summing 
over the component stars. The resulting
emission histories are quite distinct, with extended late-time flux when
lower mass components are present.
  \label{fbn_photons}}
\end{figure}

Since we are considering only non-interacting cases, their resulting UV flux can easily be obtained by 
summing over the individual contributions from the two component stars. Figure~\ref{fbn_photons} shows the ionizing 
photon flux of the S45+S15, S30+S30 and S60 models, respectively, and Table~\ref{ion.binary} lists the total 
number of ionizing photons emitted. The UV fluxes of the three models are comparable 
within a factor of 2. However, a single star, S60, still produces stronger flux than binary 
systems with the same overall mass. Differences are largest in the  
amount of ionizing \HeII{} photons, as this rate is extremely sensitive to the mass of the primary.

\section{Results}
\label{results}
The impact of the first binary stars on the early universe can be divided 
into three different classes; UV-radiation, supernova, and x-ray. 
The UV-radiative feedback here refers to the soft (LW) and hard (ionizing) photons produced
by the binaries during their stellar evolution. To facilitate
convenient comparison, we also briefly discuss the feedback from single Pop~III stars,
referring the reader to the extensive literature on this topic for further detail.
We note that our single star models improve on earlier treatments by including some
key features, such as the realistic modeling of the time dependence of the UV fluxes
in response to the underlying stellar evolution.
We present the results from our cosmological simulations following the 
chronologically order of how the first single or binary stars evolve: 
birth, evolution, and demise. { Because the results contain many cosmological simulations
of different feedback models, we summarize the stellar models, their feedbacks, and associated results in Table~\ref{crossref}.}

\subsection{Radiative Feedback}
\label{flms_results}
The first stellar system forms inside a minihalo, with a total mass of about 
$10^6\,\Msun$ and a virial radius of $\sim 100$\,pc, located in the
region with the highest mass resolution at $z \sim 27$. This allows us to resolve
key small-scale features of the ensuing stellar feedback. Once the gas density
inside the star-forming cloud exceeds the threshold for sink creation, either
a single star or a binary system is assumed to promptly form.
By using sink particles, the realistic assembly history of a protostar via an
extended phase of accretion is not modelled, which still is computationally
prohibitive. Instead, we assume that the sink particles immediately represent
fully developed Pop~III stars, acting as sources of UV radiation. The
gas inside the halo is rapidly photo-heated up to temperatures of $T\approx 2\times 
10^4\,\K$. This drives the sound speed up to $30\,\kms$, whereas the escape 
velocity of the host halo is only about $3\,\kms$. The photo-heated gas is thus blown
out of the shallow potential well of the minihalo. 
UV photons not only heat up the gas but also
ionize the neutral hydrogen and helium. The ionization-front (I-front) propagation begins with a 
short supersonic phase (R-type), then switches to a subsonic phase (D-type), because the 
I-front is trapped behind a hydrodynamical shock \cite[e.g.][]{gb01, whalen2004,wet10}. The I-front eventually breaks out, jumping ahead of the shock, and supersonically propagates
into the low-density IGM. 

\begin{figure}[h]
\begin{center}
\includegraphics[width=.8\columnwidth]{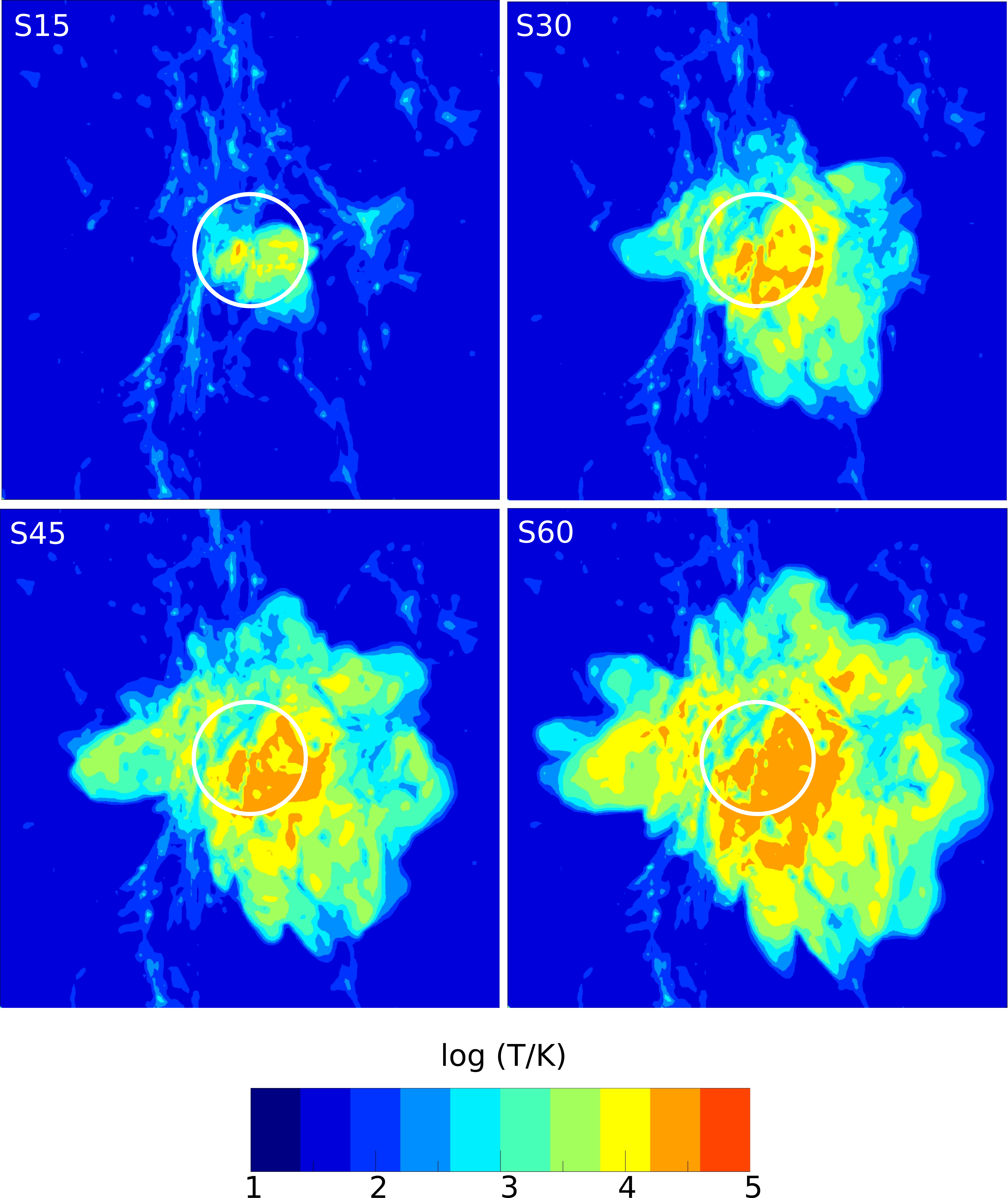} 
\end{center}
\caption{Single star radiative feedback.
The 2D maps show the gas temperature around the first 
stars right after they die. The white circle has a radius of $1\,\kpc$, 
and its center is located at the position of the first star.  \label{flms_2d_rad}}
\end{figure}

Since the lifetime of the individual stars is different, we compare their radiative feedback 
when they die and their UV radiation is terminated. We first show the resulting gas temperatures 
around the host minihalo in Figure~\ref{flms_2d_rad}. A giant bubble of hot, ionized, gas is created around the central star, with an inner
region that has reached temperatures in excess of $10^4\,\K$.
The shapes of these bubbles are very irregular due to the inhomogeneous and anisotropic 
distribution of the gas in the surrounding IGM. The bubble sizes reflect the strength of the UV emission rates, which highly depend on stellar mass.
Specifically, our S60 model creates the largest bubble with a size of
about $5\,\kpc$. The S15 model, on the other hand, only gives rise to an ionized region 
with a size of $\sim 2\,\kpc$, and a much cooler gas temperature.

\begin{figure}[h]
\begin{center}
\includegraphics[width=\columnwidth]{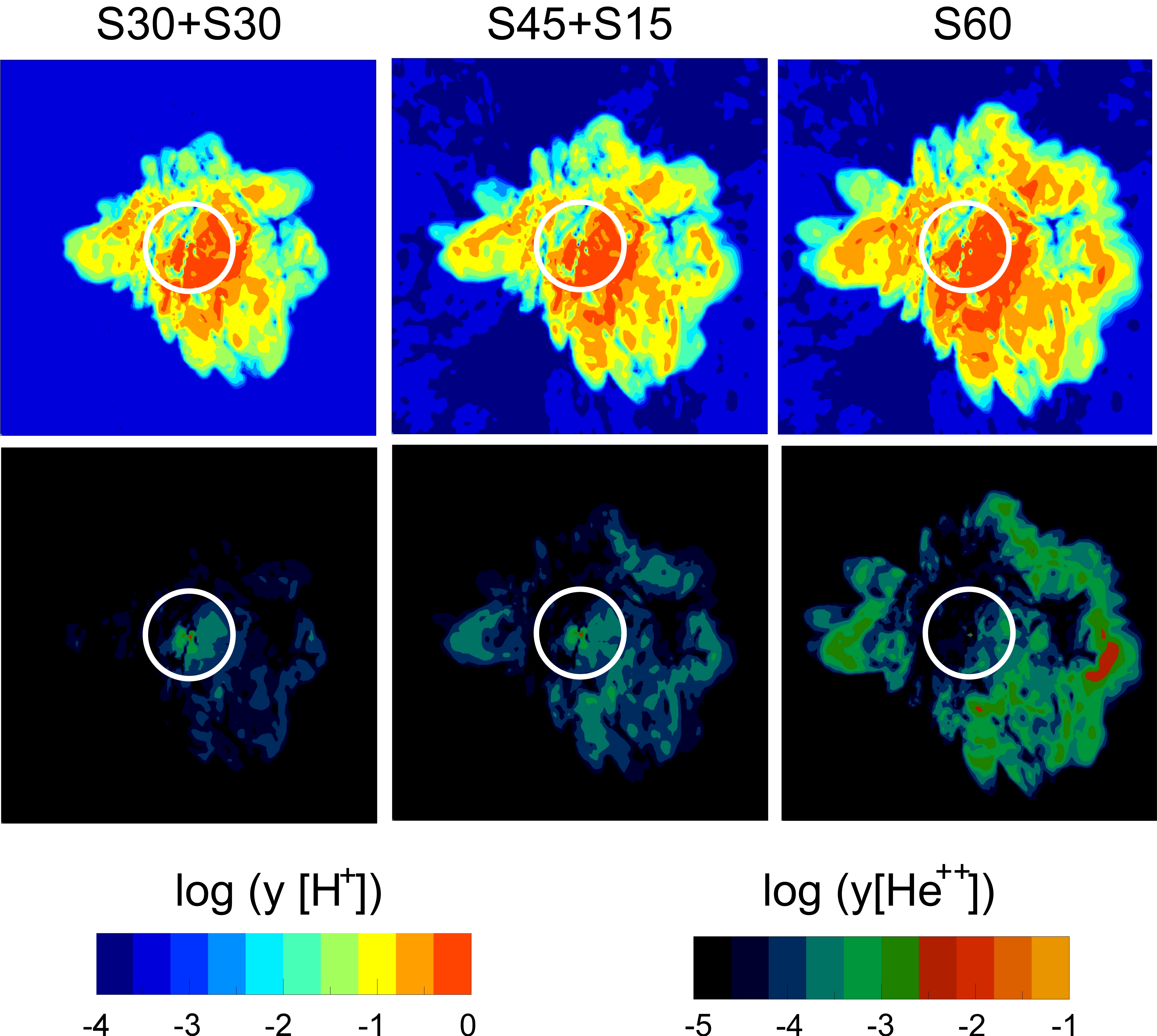} 
\end{center}
\caption{Radiative feedback from single and binary stars. 
The white circle has a radius of $1\,\kpc$, and its center is located 
at the position of the first star/binary. 
The 2D maps show the H$^+$ and He$^{++}$ regions,
right after the stars die. The size of the H$^+$ region is comparable 
among the three models. But the He$^{++}$ region is much larger in the S60 case.  
\label{fbn_2d_rad}}
\end{figure}

For binary stars, we compare cases of equal total stellar mass in
Figure~\ref{fbn_2d_rad}. Overall bubble sizes for the equal-mass models are comparable. 
However, there is a significant difference in the resulting He$^{++}$ regions.  Ionizing  He$^{+}$ 
requires photon energies of $h\nu > 54.4$\,eV, four times higher than the threshold
to ionize neutral hydrogen. Here, it greatly matters how the available stellar
mass is divided among the individual components, such that the S60 model exhibits
significantly larger He$^+$ ionizing rates than the binary models of equal mass,
S30+S30, and S45+S15. This difference may be reflected in the strength of the He$^{+}$ recombination line at $1640\,\textup{\AA}$, providing a
distinctive signature to distinguish between Pop~III single and binary systems, since the latter create 
much smaller He$^{++}$ regions, for a fixed total stellar mass.
 
\begin{figure}[h]
\begin{center}
\includegraphics[width=\columnwidth]{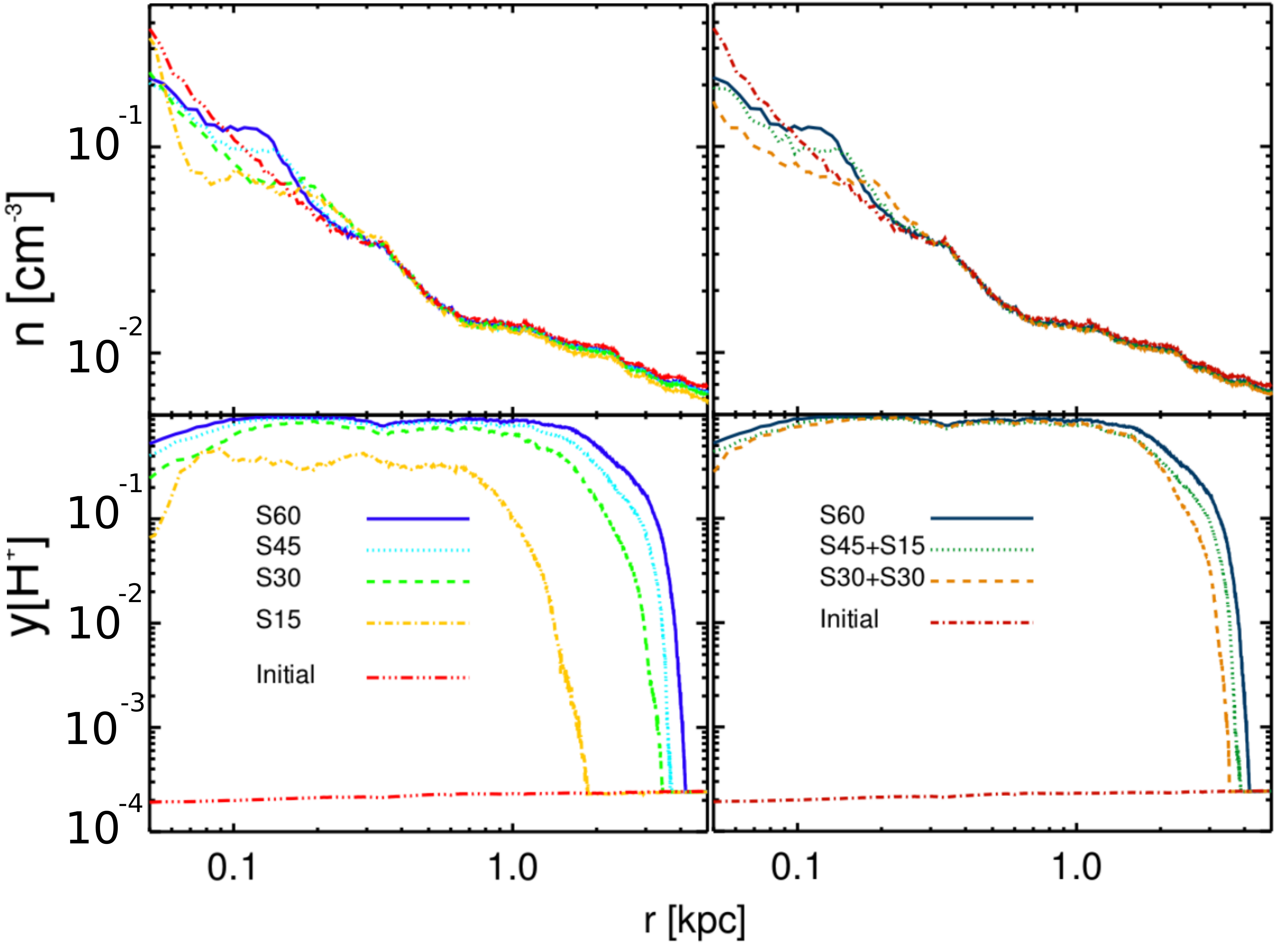} 
\end{center}
\caption{Density and ionization structure around the first star. 
The curves are mass-weighted profiles of gas density and the H$^+$ fraction inside 
the hot gas region created by the UV radiation. The red lines represent the 
conditions right before the birth of stars. There is a quick drop in y[H$^+$], the fraction 
of ionized hydrogen,
showing the boundaries between the photo-heated and unheated gas.  \label{flms_1d_h2}}
\end{figure}

To better evaluate the impact of the radiative feedback, we map the 3D structure of the hot bubbles onto 
1D radial profiles in Figure~\ref{flms_1d_h2}. We first discuss the gas density profile. Due to the 
UV photoheating, the gas density in the center of the minihalo has dropped to $0.2\,\cc$ at the end of star's lifetime. The baryonic outflow 
extends to a radius of $150\sim200~\pc$, slightly larger than the virial radius of the host halo, such that any subsequent star formation is suppressed by expelling the gas through
this photo-evaporation.
Besides their hydrodynamic feedback, UV photons also affect the chemistry of the
primordial gas by changing its ionization state, and releasing free electrons which
can catalyze H$_2$ formation. The weaker UV emission of the S15 model results in a
relatively smaller H$^{+}$ region, whereas those of
S60, S45, and S30 have radii close to $2\,\kpc$. 
The difference in the central gas density profile is mainly due to the duration over which photo-heating is active. 
Binary models exhibit longer overall lifetimes, allowing the gas to escape farther into the IGM, such that the resulting gas densities within the halo are lower.
It is not clear whether the UV radiation 
can penetrate into nearby minihalos and affect their star formation or not. When the stars die, and if there 
are no additional heating sources, the ionized gas will cool and then recombine, eventually extinguishing the fossil H$^{+}$ regions. The relevant timescales can be estimated as follows \citep{bromm2002}. The cooling time of primordial gas is approximately
$t_{\rm cool}\approx nk_{\rm B}T/\Lambda 
\approx 10^5\sim 10^6\,\yr$, where $\Lambda \propto n^2$ is the cooling rate, and $k_{\rm B}$
the Boltzmann constant. For the
recombination timescale, we estimate $t_{\rm rec}\approx (k_{rec} n)^{-1}\approx
10^6\sim 10^7\,\yr$, where the recombination coefficient is $k_{\rm rec} \approx 10^{-12}\,\cm^3 {\rm s}^{-1}$, 
and the IGM densities
$n\approx 0.01\,\cm^{-3}$. 

\subsection{Supernova Feedback}
The majority of Pop~III stars may finally die as supernovae or directly collapse 
into black holes. In this section, we discuss the supernova feedback from single 
and binary stars. When the stars die, we assume for simplicity that their  
UV radiation is immediately shut off. We employ the SN explosion energies and
metal yields discussed in Section~3, depending on the properties of the progenitor 
star. The SN explosion creates a strong
shock wave, traveling with a velocity of $v_{\rm sn} \approx 10^{4}\,
\kms$. The energy carried by the shock is able to reheat the relic \HII{} region 
for an additional $t_{\rm sn} \approx r_{h}/V_{\rm sn}\approx 0.4\,\Myr$, assuming
\HII{} region radii of $r_{h}\approx 4\,\kpc$. The shock heating in the simulation is 
roughly about $0.6\,\Myr$ because the shock velocity is slowed down due to radiative 
cooling.  After the shock dissipates, a hot and metal-rich bubble is left behind in the IGM. 
This bubble continues to expand for about another 5 million years, with an increasingly 
ill-defined boundary. Eventually, 
the thermal energy of the initial ejecta is radiated away, and the expansion stalls. The
mixing of the metals with primordial gas predominantly occurs before stalling.  
On the other hand, if a single star directly dies
as a black hole, no feedback is taken into account (but see Section~4.3). 

\begin{figure}[h]
\begin{center}
\includegraphics[width=.7\columnwidth]{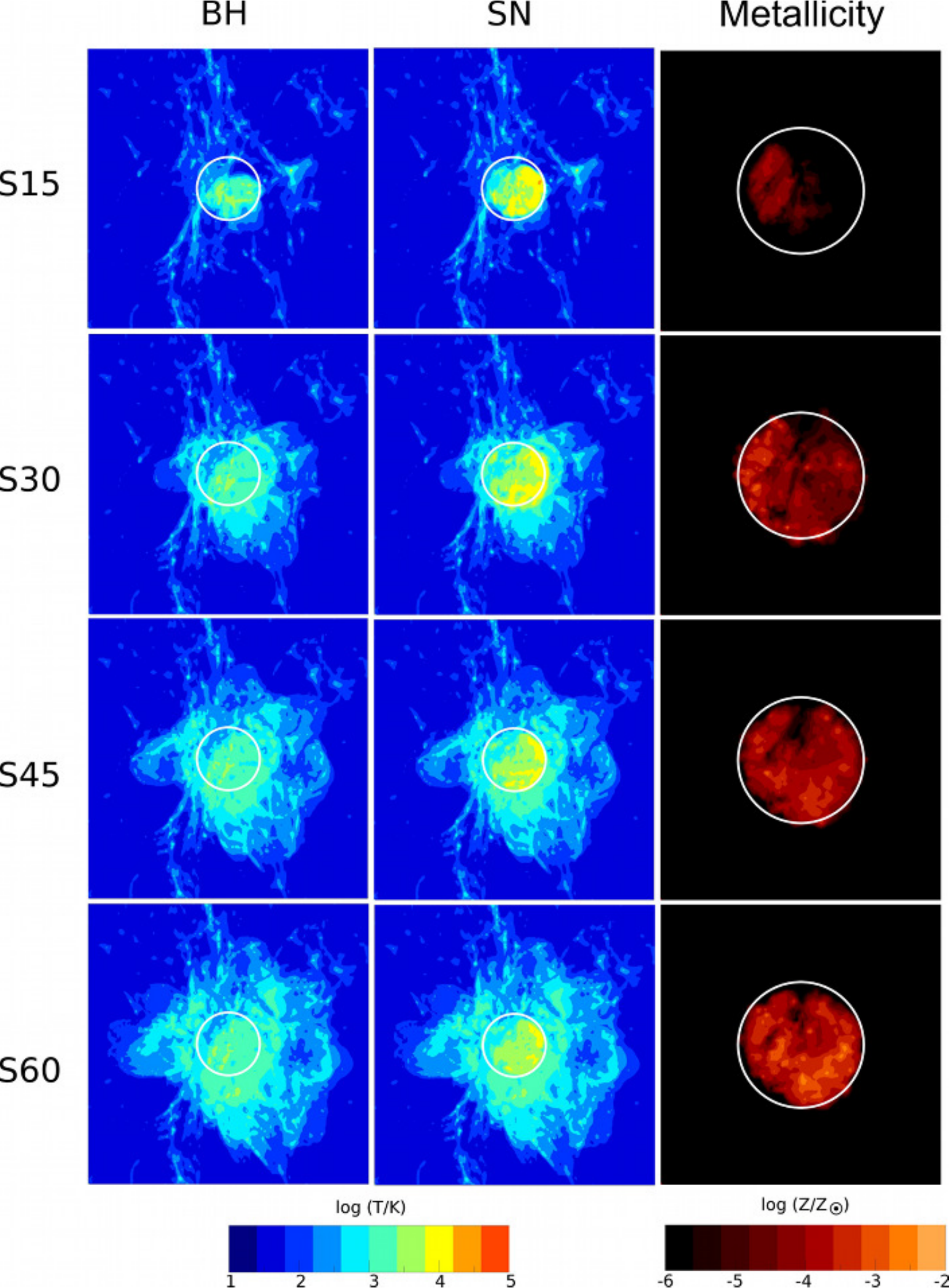} 
\end{center}
\caption{Comparison of the overall stellar feedback from all single stars 
at $15\,\Myr$ after formation. The white circle has a radius of $1\,\kpc$, 
and its center is located at the position of the first star. 
The SN ejecta stall at this time.
If the stars die as SNe, the third-column panels show the resulting metallicity distribution. The 
white circles have the same meaning as in the previous figures. In the BH scenario, 
any radiation is simply shut down, and no further stellar 
feedback occurs. Without additional heating sources present,
temperatures are significantly colder in the BH cases compared to the SN ones. 
The hypernova explosions experienced by the S30, S45, and S60 stars, disperse the metals
to $\sim 1\,\kpc$, a much larger extent than what is achieved by the S15 star.  
\label{flms_2d_all}}
\end{figure}

\begin{figure}[h]
\begin{center}
\includegraphics[width=.8\columnwidth]{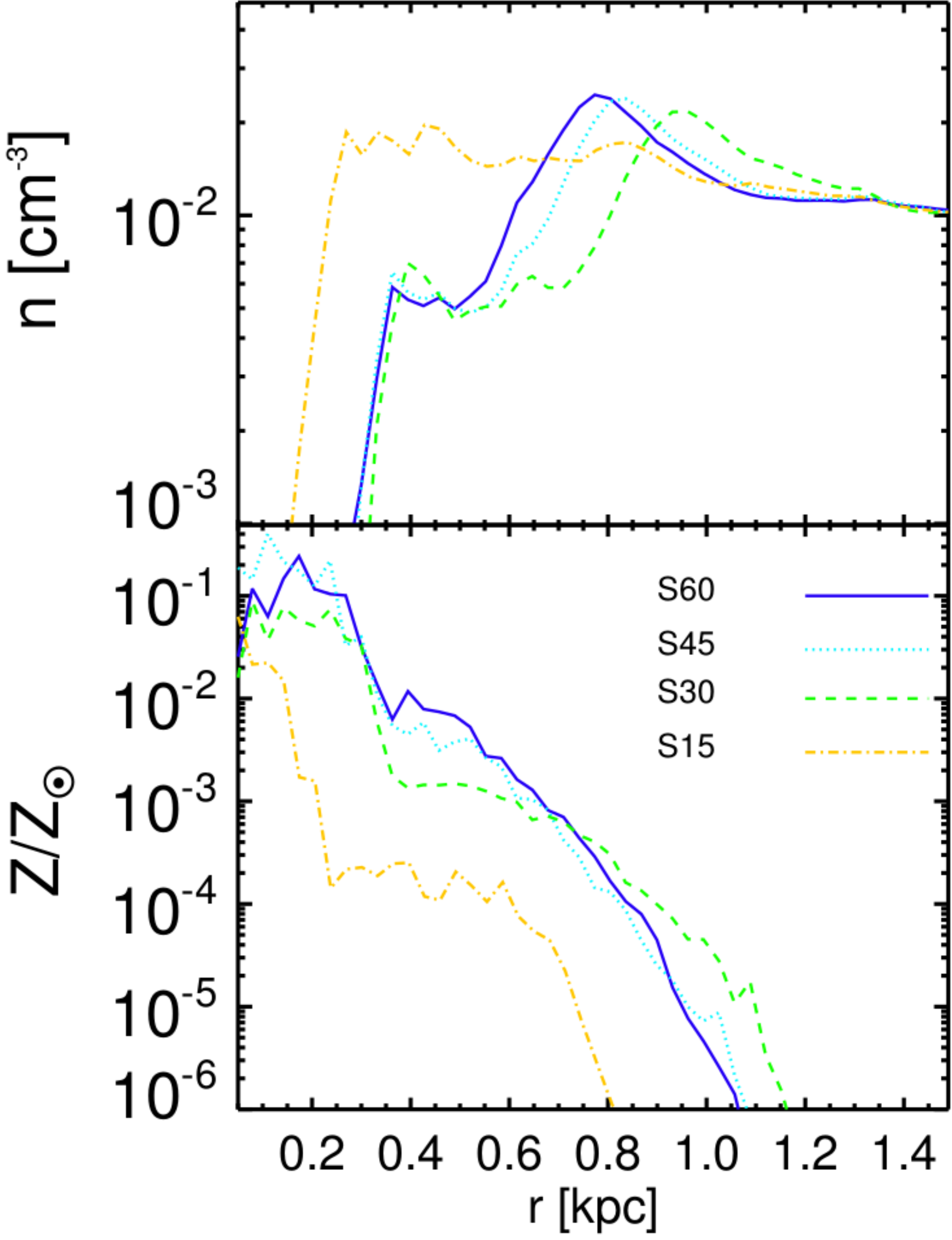} 
\end{center}
\caption{Strength of SN feedback from single stars. Shown are 1D gas density and
metallicity profiles.
The gas density inside the host halos has dropped to less than $0.01\,\cm^{-3}$. 
The ejected metal is mixed out to a radius of $\sim 1\,\kpc$
for S60, S45, S30; and $\sim 0.6\,\kpc$ for S15.
\label{flms_1d_metal}}
\end{figure}

We first discuss the combined feedback from UV radiation and supernovae/black holes,
evaluated $15\,\Myr$ after the birth of the stars, in Figure~\ref{flms_2d_all}.
At this moment, the SN blastwave has stalled. The heating from the hot SN ejecta
maintains elevated ionization in the central part of the \HII{} region. The chemical 
feedback of the S15 model is initiated by a CCSN with explosion energy of about
$1.2\,\B$ and an ejected metal mass of 
$1.4\,\Msun$. These metals are dispersed out to a radius of $\sim 0.5\,\kpc$, resulting 
in an average metallicity of $\sim 10^{-5}-10^{-4}\,\Zsun$. According to our stellar
model assumptions, the S30, S45, and S60 cases all die as 
a hypernova, dispersing 6.9, 13.3, and 20.7 $\Msun$ of heavy elements, respectively. 
For the hypernova cases, the metal-enriched bubbles
have reached radial sizes of $\sim 1\,\kpc$ with corresponding metallicities
of $10^{-4}-10^{-2}\,\Zsun$. Although all hypernova cases exhibit a similar overall range
of chemical enrichment, there are noticeable differences in the detailed distribution
of metals. These differences in turn may reflect how the preceeding radiative feedback from the
different model stars has shaped the gas distribution inside the IGM, which could
affect the transport of metals later on. 
To further facilitate the comparison between the single-star cases, we present
1D profiles of gas density and metallicity in Figure~\ref{flms_1d_metal}.  
The strong SN blast waves in all cases substantially suppress the gas density in the
host halo, resulting in $n < 0.01\,\gcc$, similar to that of the background IGM at this
redshift of $z \sim 26$.
The extent of the central void is about $200\,\pc$ for the CCSN
case (S15), and $500\,\pc$ for the HN cases, respectively.
The metallicity profiles for the HN cases are quite similar, but the S30 model
enriches the IGM out to a slightly larger radius in response to its higher
blast wave velocity. Finally, instead of
SN explosions, the stars may directly collapse into a BH, a possibility
which becomes increasingly likely with increasing progenitor mass.
For such a BH fate, the relic \HII{} regions would quickly begin to cool to 
about $10^3\,\K$, and recombination would suppress the abundance of free electrons.

\begin{figure}[h]
\begin{center}
\includegraphics[width=.8\columnwidth]{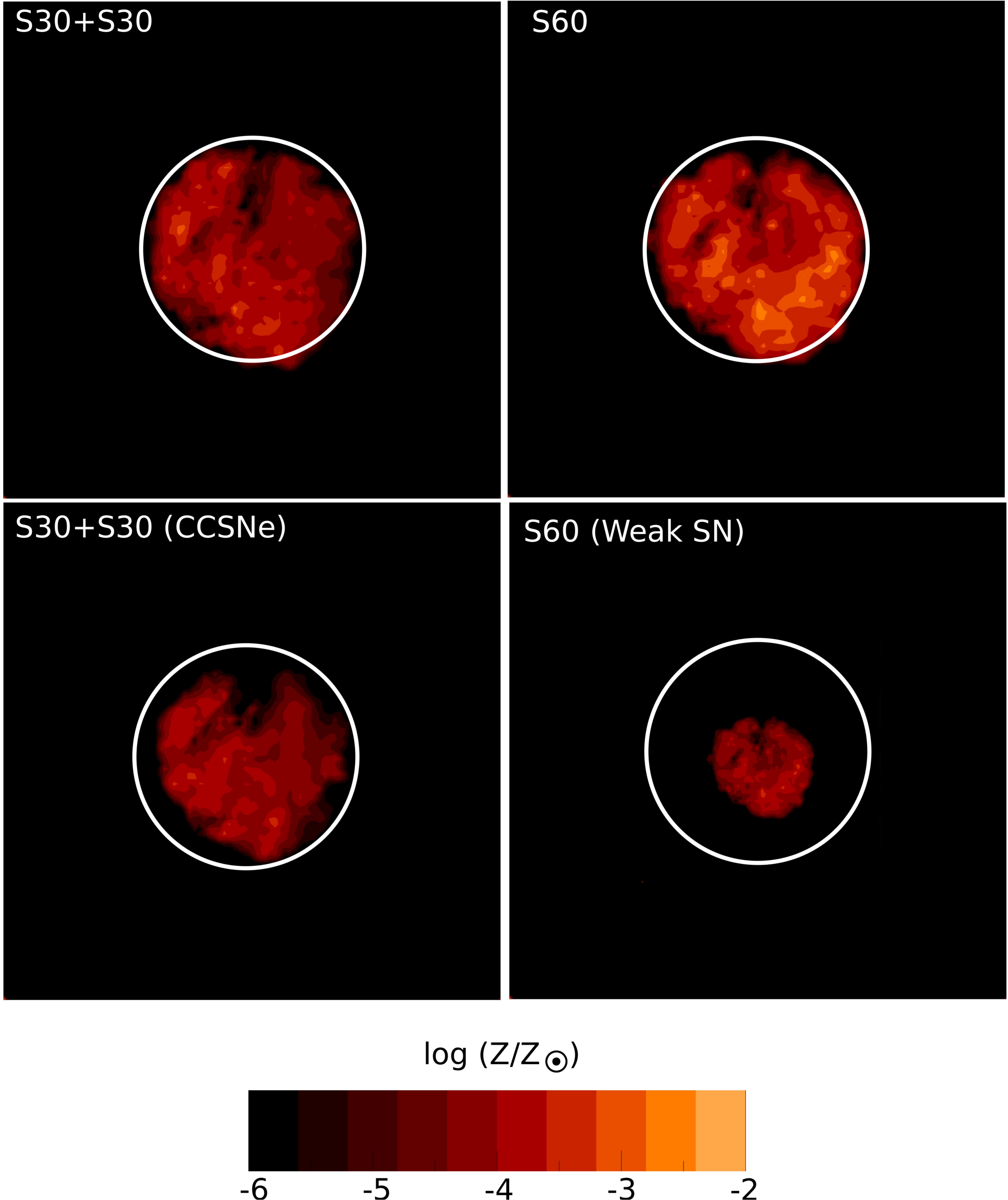} 
\end{center}
\caption{Chemical enrichment from the first binaries. The white circle has a radius of $1\,\kpc$, 
and its center is located at the position of the first star/binary. We here compare the two 
different scenarios for the
S30+S30 and S60 models. For S30+S30, both stars can die as hypernovae or core-collapse 
supernovae. In S60 (Weak SN), we show an example of a weak explosion by scaling down
the explosion energy and metal yield of a hypernova 
by a factor of 100. In a weak explosion, 
most of the metals fall back onto the BH without being ejected.\label{fbn_metal}}
\end{figure}

\begin{figure}[h]
\begin{center}
\includegraphics[width=.8\columnwidth]{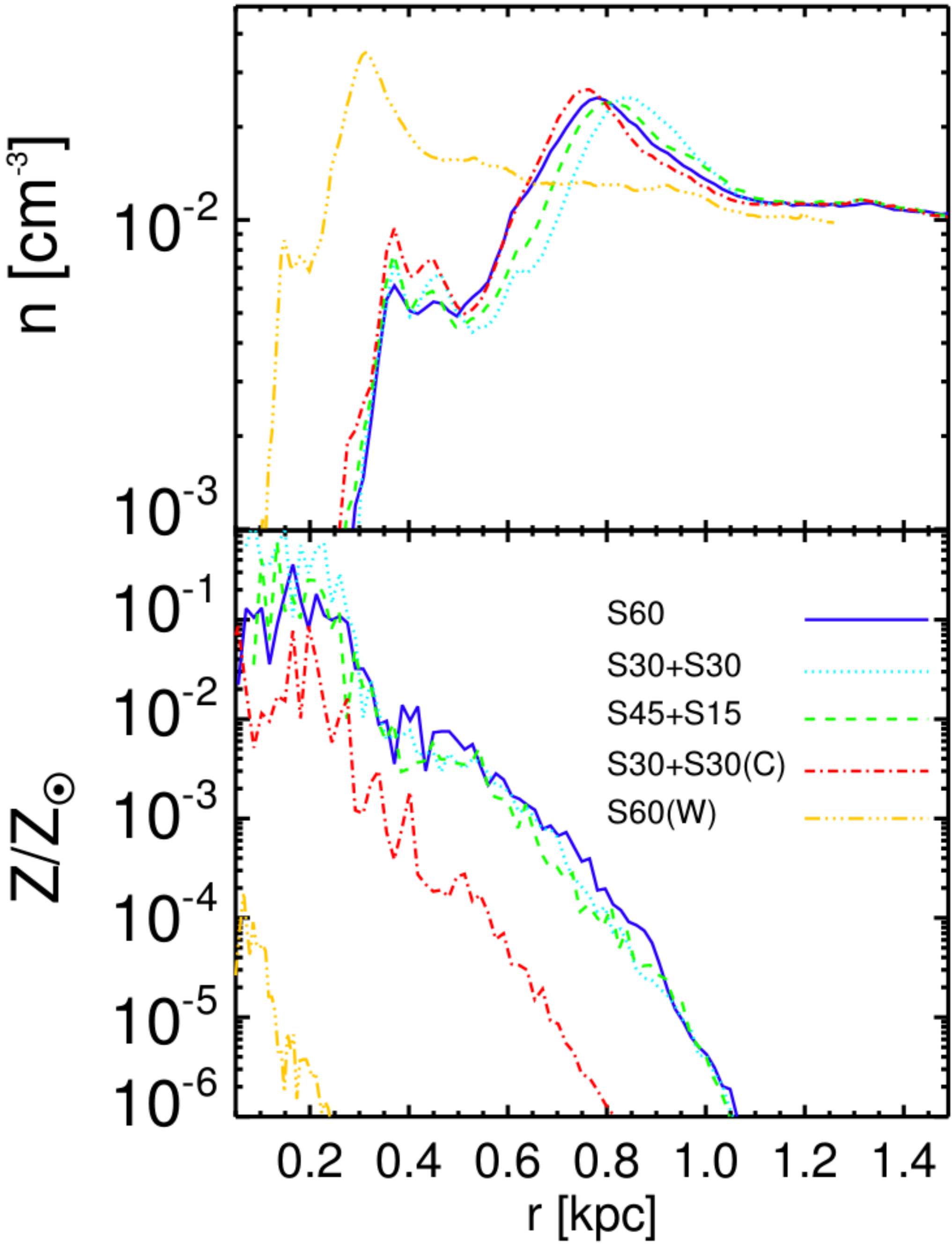} 
\end{center}
\caption{SN feedback from single and binary stars. The two panels
show 1D gas density and metallicity profiles inside the SN 
ejecta. S60, S30+S30, S45+S15 are hypernova explosions. 
S30+S30(C) is a core-collapse SN, and S60(W) is a weak SN. 
\label{fbn_1d_metal}}
\end{figure}

We now discuss the SN feedback from binary stars. 
To enable a meaningful comparison, we fix the total mass in Pop~III
stars, here $60\,\Msun$, either locked up in a single star or distributed among
binary partners (S30+S30 or S45+S15). The key question then is whether there are
significant differences or not.
Stellar evolution models strongly
suggest that $60-80\,\Msun$ Pop~III stars are likely to die as BHs, possibly
accompanied by very weak SN explosions.
Hence, we also consider the case of a $60\,\Msun$ star dying as such a weak 
SN with an explosion energy of $\sim 0.1 $B and a metal yield of
$\sim 0.1 \Msun$, due to strong fallback during the BH-forming explosion.
The basic trends can be gleaned from   Figure~\ref{fbn_metal} and Figure~\ref{fbn_1d_metal}. 
As long as at least one component explodes as a HN, the resulting
metal enrichment is very similar. However, the enrichment from the binary systems is
more robust, in the sense that the single S60 star is likely to experience only
a weak explosion accompanied by much reduced heavy element production and dispersal.
Thus, the recent revision of the Pop~III star formation paradigm away from a single-star
outcome to ubiquitous binarity in effect enhances chemical feedback in the early
universe.

 \subsection{X-Ray Feedback}
\begin{figure}[h]
\begin{center}
\includegraphics[width=\columnwidth]{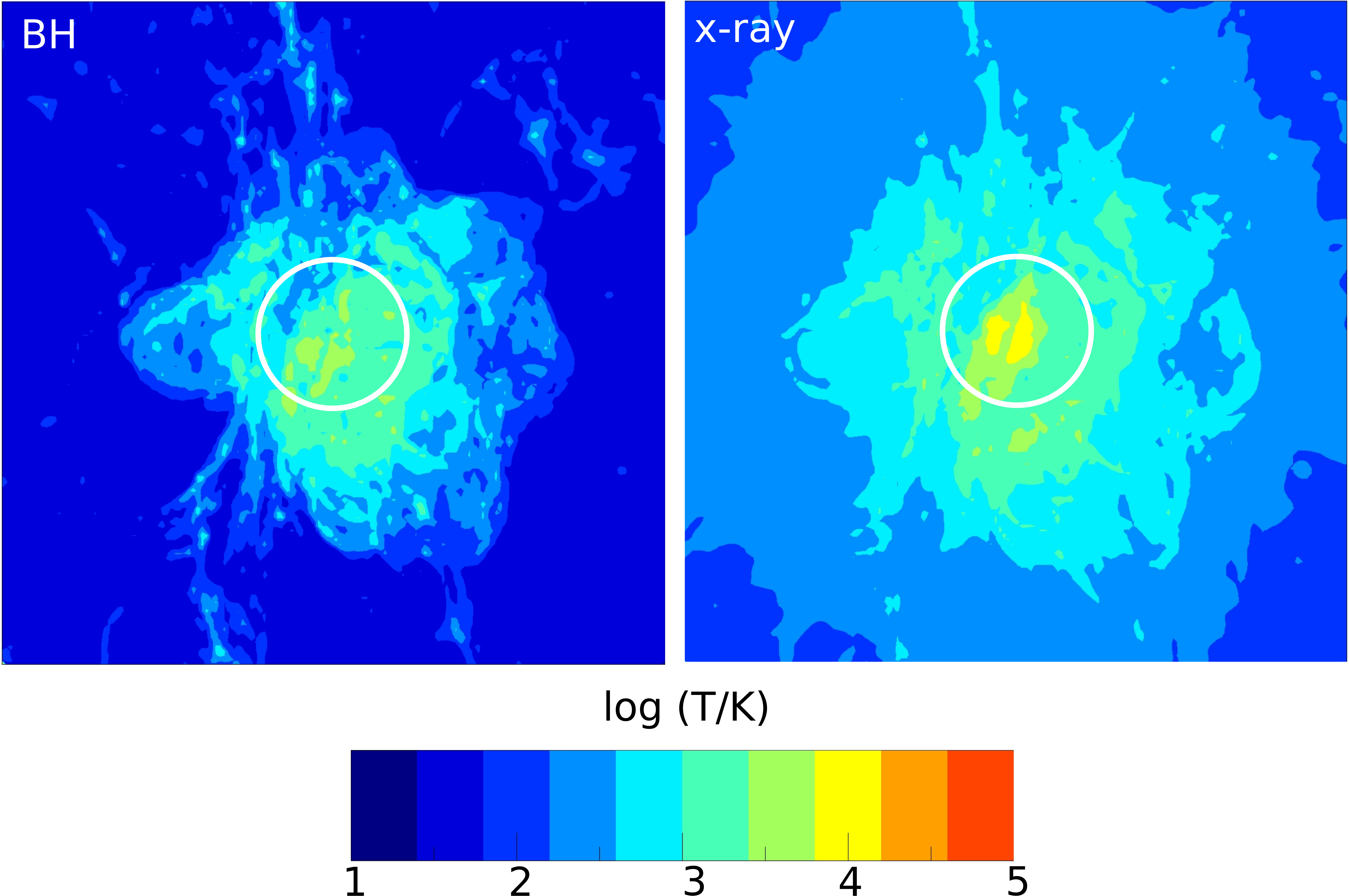} 
\end{center}
\caption{Radiative feedback from a Pop~III x-ray binary. 
The white circle has a radius of $1\,\kpc$, 
and its center is located at the position of the first star. 
{ The snapshots are taken at the time about 15 \Myr{} after the 
binary formation}.
The 2D colored map shows the temperature without and with the 
x-ray binary. 
The x-ray emission heats 
up the relic {\HII} region and beyond. The morphology of the
feedback is rather isotropic and homogeneous.
 \label{fbn_xray}}
\end{figure}

 One particularly interesting variant of the S45+S15 model is the presence of
 long-lived x-ray feedback.
 Because of the uncertainty in the stellar evolution model, the $45\,\Msun$ star
 can possibly collapse into a BH instead of blowing up as a SN. In the case of 
 a close binary, mass transfer from the $15\,\Msun$ companion, still alive at this time,
 becomes possible. Mass accretion onto the compact object
 can efficiently extract the gravitational energy of the infalling material, converting
 it into the thermal energy within the accretion disk that leads to x-ray 
 emission. Unlike the ionizing photons from stars, the x-rays can more easily penetrate 
 the IGM because of the much reduced opacity at high energies \cite[e.g.][]{mba03,kuhlen2005, jeon2012}. 
 Here, we assume a constant accretion rate onto the central BH of about 
 $10^{-6}\,\Msun\,\yr^{-1}$, which lasts for about 10 million years. 
 For simplicity, we further assume that the
 $15\,\Msun$ companion star eventually dies as a white dwarf without a SN explosion, and
 neglect its UV radiation during the accretion phase. We 
 show the impact of such a Pop~III x-ray binary, comparing it with the non x-ray case, in 
 Figure~\ref{fbn_xray}. In the absence of x-ray emission, the temperature of the
 the relic \HII{} region quickly declines and the ionized hydrogen recombines.
 An active x-ray binary, on the other hand, provides a prolonged heating source,
 maintaining the temperature and ionization inside the relic \HII{} region, and further 
 heating up the gas beyond its boundary, up to several hundred 
 Kelvin. More importantly, the x-ray emission can change the free electron fraction
 in the IGM which is critical for 
 H$_2$ formation.
 Because the IGM is optically thin to x-ray photons, the resulting heating is 
 quite homogeneous and isotropic. We also find that the x-ray photons may penetrate the 
 gas of nearby halos, thus affecting their star formation properties.

 \section{Discussions and Conclusions}
 \label{flms_discussion}
 We have presented the results from cosmological simulations of the 
 impact of Pop~III stars, specifically focusing on the new effects arising from the 
 presence of binaries. We improve on earlier simulations by using
 updated Pop~III stellar models. In ascertaining the cosmological impact of
 the first binaries, we consider their radiative, SN, and x-ray feedback. 
 By comparing a single $60\,\Msun$ star and the corresponding 
 binary systems with equal total mass, we find that the resulting numbers of
 hydrogen-ionizing photons are very similar. 
 However, for He$^+$ ionizing radiation, the binary stars are significantly 
 weaker than the single star, because the more energetic UV photon production
 is strongly reduced
 in the less massive stars. If binary stars thus were the typical outcome of
 Pop~III star formation in minihalos, detection of
 the distinct emission lines from He$^{++}$ recombination, most prominently the
 1640 $\textup{\AA}$ line, would be very challenging. 
  {Because the x-ray feedback strongly depends on the spectrum of 
  x-ray binaries which is uncertain in the high redshifts, its observational signature 
  is more difficult to predict. Recently, \citet{xu2014} and \cite{ahn2014}  suggested 
 that the x-ray feedback of the Pop III binaries can be examined by the 21 cm observations.}

 We here explore cases where the stars die as core-collapse SNe or as 
 hypernovae. In all cases, the mechanical feedback from the explosions  
 expels the gas from the host systems, thus suppressing any subsequent
 star formation in the same halo, at least for of order 10\,Myr. To trigger any
 further star formation, the expelled gas needs to be driven back to high 
 density, possibly as a result of halo mergers within bottom-up structure
 formation later on.
 For a $15\,\Msun$ star, the expected final fate is a core-collapse  SN, whereas
 the fate of $30\,\Msun$, 
 $45\,\Msun$, and $60\,\Msun$ Pop~III stars is still poorly understood. To
 bracket parameter space, we here assume
 that they die as energetic hypernovae, weak SNe, or directly collapse into BHs.
 It is evident that for such more massive progenitors, chemical feedback
 can span a broad range of possibilities. The most effective feedback is 
 provided by the hypernova models, where the nearby IGM is typically enriched to
 average metallicities of $\sim 10^{-4} - 10^{-3}\,\Zsun$, out to
 $\sim 2\,\kpc$.  
 
\begin{figure}[h]
 \begin{center}
 \includegraphics[width=.7\columnwidth]{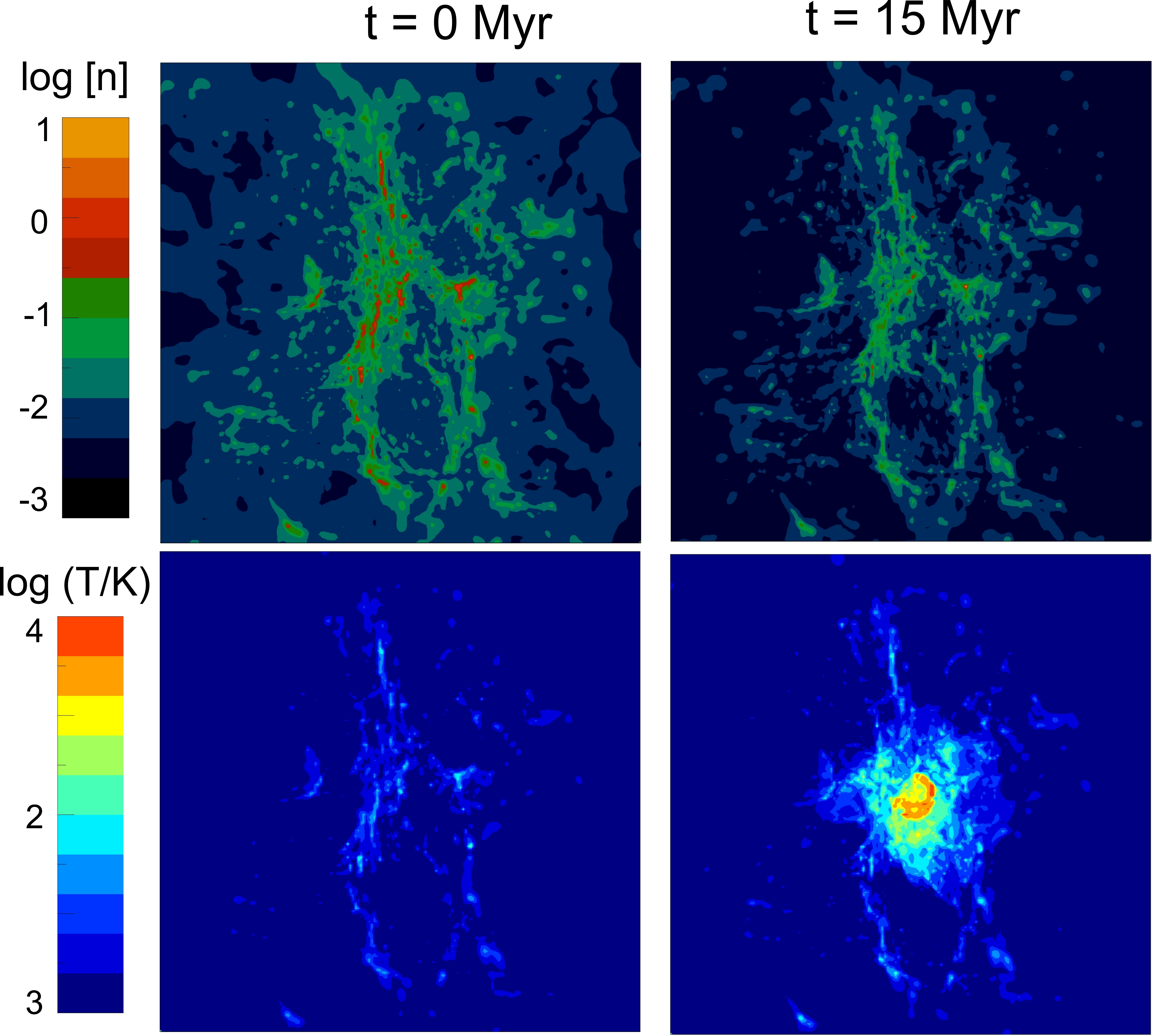}
 \caption[]{Cosmological impact of a $60\,\Msun$ star. 
 Panels show temperatures and densities before/after the stellar feedback from a 
 $60\,\Msun$ star. { The left two panels are at t= 0 , right before the star formation 
 and the right two panels are at t= 15 Myr}. The size of each panel is about $10\times10\,\kpc^2$. Some 
 higher density clumps have been smoothed due to the radiative and SN feedback which 
 also chemically enriches the pristine gas within the orange circle of radius
 $\sim 1\,\kpc$. Both radiative and supernova feedback heat up the gas and change its 
 chemistry on a scale of $3-4\,\kpc$. \label{impacts_den_temp}}
 \end{center}
 \end{figure}

 Even single, energetic SNe can impact the early universe on cosmological scales.
 We demonstrate this in Figure~\ref{impacts_den_temp}, showing the feedback from 
 a single $60\,\Msun$ star, undergoing a HN explosion.
 The resulting feedback significantly enhances the IGM temperature, smoothes out
 density structures in nearby halos, and enriches the primordial gas over regions 
 of $\sim 2\,\kpc$. However, the formation of a single massive star inside a
 minihalo increases the probability of collapsing into a BH without any 
 chemical enrichment. On the other hand, binary star formation greatly buttresses
 the likelihood that metal enrichment will occur. { In effect, binary formation 
 is much less likely to keep the early universe metal free, with consequences for the
  prompt transition in star formation mode to the low-mass dominated Population~II.}

 Realistic binary stellar models could introduce a very rich phenomenology of
 evolutionary pathways. Among them are ejection scenarios, where one component is
 flung out, such that it may explode in the outskirts of its host halo, as opposed
 to the location of its birth, as implicitly assumed here. There is also
 rich physics related to the early evolution of the SN remnant, when shock
 break-out occurs, and hydrodynamical mixing takes place. All of these
 intriguing aspects of binary progenitors will be explored in future simulations,
 with greatly improved spatial and temporal resolution. The exploratory models
 presented here, however, already clearly indicate the importance of studying
 binary-related feedback in the early universe. The imprint from the 
 violent death of Pop~III stars, in all its variety, might soon be amenable to
 empirical testing with the {\it James Webb Space Telescope (JWST)}, to be launched
 around 2018. One key aspect of this search will be to distinguish the possible signature
 of binarity in primordial star formation.

 \section*{Acknowledgments}
 {Authors thank the anonymous referee for many constructive suggestions}.
 We also thank Jarrett Johnson, Dan Whalen and Ryan Cooke for many useful discussions.
 KC was supported by a IAU-Gruber Fellowship, a Stanwood 
 Johnston Fellowship, and a KITP Graduate Fellowship. Work at UCSC has been 
 supported by DOE HEP Program under contract DE-SC0010676; the NSF 
 (AST 0909129) and the NASA Theory Program  (NNX14AH34G).
 VB acknowledges support from NSF grant AST-1009928 and NASA grant NNX09AJ33G. 
 AH acknowledges support by an ARC Future Fellowship (FT120100363) and a 
 Monash University Larkins Fellowship. This work has been supported by the DOE 
 grants; DE-GF02-87ER40328, DE-FC02-09ER41618 and by the NSF grants; 
 AST-1109394, and PHY02-16783. All numerical simulations were performed with allocations from the University 
 of Minnesota Supercomputing Institute and the National Energy Research Scientific Computing Center.

\newpage

\begin{table}[h]
\begin{center}
\begin{tabular}{ccccccc}
Model &
Mass &
MS &
post-MS &
total &
fates &
metal yield 
\\ \hline
{} &
($\Msun$) &
($\Myr$) &
($\Myr$) &
($\Myr$) &
{} &
($\Msun$) 
\\ \hline
 S15 & $15$ & $9.478$ & $1.031$ & $10.51$ & SN     & $1.388$ \\
 S30 & $30$ & $5.208$ & $0.509$ & $ 5.77$ & BH, HN & $6.876$ \\
 S45 & $45$ & $3.995$ & $0.394$ & $ 4.39$ & BH, HN & $13.26$ \\
 S60 & $60$ & $3.426$ & $0.345$ & $ 3.77$ & BH, HN & $20.66$  
 \\  \hline
\end{tabular}
\caption{Stellar lifetimes and fates.}
\label{flms_1ife_models}
\end{center}
\end{table}

\begin{table}[h]
\begin{center}
\begin{tabular}{lclll}
{Type} &
{Masses} &
{$E_{\rm sn}$} &
{mass ejection} &
{Notes}
\\ \hline
{} &
{($\Msun$)} &
{($\foe$)} &
{} &
{} 
\\ \hline
 SN & $\lesssim25$ & 1.2 & all but $\sim1.5\,\Msun$ & leaves neutron star \\
BH & $\gtrsim25$ & 0   &                     None & complete collapse to BH \\
 HN & $\gtrsim25$ & 10  & $\sim90\,\%$             & energetic explosion \\  \hline
\end{tabular}
\caption[Summary of assumed stellar fate characteristics]{Summary of assumed stellar fates. $E_{\rm sn}$ is the explosion energy.}
\label{flms_fate}
\end{center}
\end{table}

\begin{table}[h]
\begin{center}
\begin{tabular}{cccccc}
\\ \hline 
Model & Mass & \HI  & \HeI & \HeII \\
{} & $(\Msun)$ & $(10^{63})$ & $(10^{63})$ & $(10^{61})$& \\
\hline
S15 & $15$      & $0.64$ & $0.16$ & $0.10$ \\
S30 & $30$      & $1.82$ & $0.72$ & $1.37$ \\
S45 & $45$      & $2.98$ & $1.45$ & $4.34$  \\
S60 & $60$      & $4.18$ & $2.21$ & $8.31$  \\
\hline
\end{tabular}
\caption{Number of ionizing photons emitted over the lifetime of a star.
\label{flms_flux}}
\end{center}
\end{table}

\begin{table}[h]
\begin{center}
\begin{tabular}{cccccc}
\\ \hline
{Case} &
{Masses} &
{Separation} &
{Fate} &
{Fate} &
metals yields
\\ \hline
{}    &
{($\Msun$)}    &
{(distance)}  &
{1} &
{2} &
($\Msun$) 
\\ \hline
I    & 30+30 & wide  & HN & HN & $13.74$\\
II    & 30+30 & wide  & BH & BH & $0.00$\\
III  & 45+15 & close & BH & WD  & $0.00$ \\
 \hline
\end{tabular}
\caption[Summary of binary model characteristics]{Summary of binary model characteristics.}
\label{binaryfate}
\end{center}
\end{table}

\begin{table}[h]
\begin{center}
\begin{tabular}{lcccc}
\hline 
Model & \HI{} & \HeI{} & \HeII{}  &  $t_{*}^a$ \\
& $(10^{63})$ & $(10^{63})$ &  $(10^{61})$ & $(\Myr)$  \\
\hline
S30+S30  & $3.64$ & $1.44$ & $2.74$ & $ 5.77$ \\
S45+S15  & $3.62$ & $1.61$ & $4.43$ & $10.51$ \\
S60     & $4.18$ & $2.21$ & $8.31$ & $3.77$\\
\hline
\end{tabular}
\caption[Number of ionizing photons emitted over the lifetime of binaries]{Number of ionizing photons 
emitted over the lifetime of binary models. $^a$ Lifetime of a binary star (longest-lived component).}
\label{ion.binary}
\end{center}
\end{table}

\begin{table}[H]
\begin{center}
\begin{tabular}{ |l|l|l| }
\hline
Type of star & Stellar model and feedback & Results (Figure: No)  \\ \hline
\multirow{13}{*}{Single} & S15 Rad &  3, 5\\
 & S15 Rad+CCSN &  6, 7  \\
 & S15 Rad+BH &  6 \\
 & S30 Rad &  3, 5\\
  & S30 Rad+HN &  6, 7  \\
  & S30 Rad+BH  & 6 \\
  & S45 Rad &  3, 5\\
   & S45 Rad+HN & 6, 7, 9  \\
   & S45 Rad+BH  &  6\\
  & S60 Rad &  3, 4, 5\\
    & S60 Rad+HN &  6, 7, 8, 9, 11\\
    & S60 Rad+WSN & 8, 9 \\
    & S60 Rad+BH  &  6  
      \\ \hline
\multirow{6}{*}{Binary} & S45+S15 Rad &  4, 5\\
 & S45+S15 Rad+HN &  9 \\
  & S45+S15 Rad+x-ray &  10\\
 & S30+S30 Rad & 4, 5 \\
  & S30+S30 Rad+HN & 8, 9  \\
   & S30+S30 Rad+CCSN & 8, 9 
 \\ \hline
\end{tabular}
\caption{Summary of feedback models and results. Rad presents the radiative feedback from 
the UV radiation. BH (black hole), CCSN (core-collapse SN), WSN (weak SN), HN (hypernova) are 
different fates of the star.}
\label{crossref}
\end{center}
\end{table}


\end{document}